\documentclass[aps,prd,twocolumn,superscriptaddress,tightenlines,nofootinbib]{revtex4-1}
\usepackage{geometry}
\usepackage[utf8]{inputenc}
\usepackage{graphicx}
\usepackage{amssymb}
\usepackage{textcomp}
\usepackage{amsmath}
\usepackage{tabularx}
\usepackage{bm}
\usepackage{times}
\usepackage{color}
\usepackage{slashed}
\usepackage{multirow}
\usepackage{verbatim}
\usepackage{cancel}
\usepackage{tikz}
\usepackage{epsfig,epstopdf}
\usepackage{slashed}

\usepackage[colorlinks=true, pdfstartview=FitV, linkcolor=blue, citecolor=blue, urlcolor=blue]{hyperref}
\allowdisplaybreaks[4]

\def\lsim{\mathrel{\raise.3ex\hbox{$<$\kern-.75em\lower1ex\hbox{$\sim$}}}}
\def\gsim{\mathrel{\raise.3ex\hbox{$>$\kern-.75em\lower1ex\hbox{$\sim$}}}}

\definecolor{orange}{rgb}{1,0.5,0}

\def\beq{\begin{equation}}
\def\eeq{\end{equation}}
\def\beqa{\begin{eqnarray}}
\def\eeqa{\end{eqnarray}}

\begin{document}
\DeclareGraphicsExtensions{.eps,.ps}


\title{Constraining neutrinophilic mediators at FASER$\nu$, FLArE and FASER$\nu$2}

\author{Weidong Bai}
\email{baiweidong@lut.edu.cn}

\affiliation{School of Physics, Sun Yat-Sen University, Guangzhou 510275, China}
\affiliation{Department of Physics, School of Science, Lanzhou University of Technology, Lanzhou 730050, China}

\author{Jiajun Liao}
\email{liaojiajun@mail.sysu.edu.cn}
\affiliation{School of Physics, Sun Yat-Sen University, Guangzhou 510275, China}

\author{Hongkai Liu}
\email{hliu6@bnl.gov}
\affiliation{High Energy Theory Group, Physics Department,
	Brookhaven National Laboratory, Upton, New York 11973, USA}
\affiliation{
	Physics Department, Technion – Israel Institute of Technology, Haifa 3200003, Israel
}

\begin{abstract}
 
High energy collider neutrinos have been observed for the first time by the FASER$\nu$ experiment. The detected spectrum of collider neutrinos scattering off nucleons can be used to probe neutrinophilic mediators with GeV-scale masses. We find that constraints on the pseudoscalar (axial vector) neutrinophilic mediator are close to the scalar (vector) case since they have similar cross section in the neutrino massless limit.
We perform an analysis on the measured muon spectra at FASER$\nu$, and find that the bounds on the vector mediator from the current FASER$\nu$ data are comparable to the existing bounds at $m_{Z^\prime}\approx 0.2$ GeV. We also study the sensitivities to a neutrinophilic mediator at future Forward Physics Facilities including FLArE and FASER$\nu$2 by using both the missing transverse momentum and the charge
identification information. We find that FLArE and FASER$\nu$2 can impose stronger bounds on both the scalar and vector neutrinophilic mediators than the existing bounds. 
The constraints on the scalar mediator can reach 0.08 (0.1) for $m_\phi\lesssim1$ GeV with (without) muon charge identification at FASER$\nu$2.
\end{abstract}

\maketitle
\section{Introduction}
\label{sec:Intro}
Various neutrino oscillation experiments have demonstrated that neutrinos possess nonvanishing masses that cannot be explained in the Standard Model (SM)~\cite{Workman:2022ynf}. Thus, the discovery of neutrino oscillation provides a strong motivation to search for new physics (NP) beyond the SM. In many extensions of the SM that are related to the generation of light neutrino masses, there are often predictions of a new boson that are coupled to neutrinos~\cite{Minkowski:1977sc,Yanagida:1979as,GellMann:1980vs,Glashow:1979nm,Mohapatra:1979ia,Schechter:1980gr, Gelmini:1980re, Chikashige:1980ui,Foot:1988aq}. 
The presence of a neutrinophilic boson that predominantly interacts with neutrinos rather than other SM particles often evades strong constraints in the laboratory due to the elusive nature of neutrinos, and it can also lead to the neutrino self-interactions ($\nu$SI), which are highly motivated by the neutrino mass mechanism, dark matter, and  Hubble tension; for a recent review see Ref.~\cite{Berryman:2022hds}. 
Since the neutrinophilic mediators can be produced via the bremsstrahlung off a neutrino beam during  neutrino interactions in the detector, a promising way to probe them in the laboratory is through the precision measurement of the missing transverse momentum with respect to the neutrino beam direction~\cite{Berryman:2018ogk}. 

The Forward Search ExpeRiment (FASER) is designed to study the properties of new light and weakly coupled particles and high energy collider neutrinos~\cite{Feng:2017uoz,FASER:2018eoc}, and the main spectrometer of FASER is located about 480~m from the ATLAS interaction point at the Large Hadron Collider (LHC). A dedicated FASER$\nu$~\cite{FASER:2019dxq,FASER:2021mtu} experiment that consists of a emulsion/tungsten detector is also located in front of the FASER spectrometer. FASER$\nu$ can be used to study the high-energy neutrinos produced in the forward region of the proton-proton collision at the LHC. Recently, the FASER$\nu$ experiment has made the first observation of collider neutrinos using the active electronic components of the FASER spectrometer~\cite{FASER:2023zcr}. The high-energy electron and muon neutrino cross sections have also been measured by the FASER$\nu$ emulsion/tungsten detector~\cite{FASER:2024hoe}. 
In the future, FASER and FASER$\nu$ will be upgraded to FASER2 and FASER$\nu$2, respectively. Together with other proposals such as the forward liquid argon experiment (FLArE)~\cite{Batell:2021blf}, they will consist of the Forward Physics Facility (FPF) in the high luminosity LHC (HL-LHC) era~\cite{Feng:2022inv}.

The measurement of the charged-current (CC) neutrino interactions in the unexplored TeV energy range at FASER$\nu$ provides a unique platform to probe the neutrino beamstrahlung signal along with the SM CC neutrino interaction~\cite{Berryman:2018ogk}. Neutrino beamstrahlung is a process in which the neutrino radiates a dark mediator while getting detected in a charged-current (CC) interaction. Compared to the SM CC interaction, the radiation of a dark mediator will lead to a missing transverse momentum with respect to the neutrino beam direction. Reference~\cite{Kelly:2021mcd} has studied the sensitivities to a neutrinophilic scalar at a FLArE-like detector by exploring the missing transverse momentum. In addition, the radiation of a scalar will produce a wrong sign charged lepton which will be a clean signal if the detector has the charge identification information~\cite{Barger:1981vd,Berryman:2018ogk}. The electronic components of the FASER/FASER2 detector also has the ability to reconstruct the charge of final leptons~\cite{FASER:2023zcr,FASER:2022hcn}, which can be used to improve the sensitivities to the neutrinophilic scalar.
In this work, we study the sensitivities of the neutrinophilic mediators at FPF by using both the missing transverse momentum and the charge identification. We first perform a systematical study of the cross section of neutrinos scattering on nucleons with radiation of a dark mediator that is coupled to neutrinos predominantly. 
Then we consider the constraints on the parameter space of the new scalar or vector mediator by using the current FASER$\nu$ data and from simulation of future FLArE and FASER$\nu$2 experiments. A distinctive reach in the regime of GeV-scale mass for the neutrinophilic mediators is expected from the measurement of the high energy collider neutrinos.

This paper is organized as follows. In Sec.~\ref{sec:model} we  discuss the cross section of the neutrino scattering process in the presence of a neutrinophilic mediator.
Then, we analyze the FASER$\nu$ data in Sec.~\ref{sec:analysis} .
In Sec.~\ref{sec:simulation}, we investigate the sensitivities of searching for a scalar or vector neutrinophilic mediator at future FLArE and FASER$\nu$2 experiments.
Finally, we draw our main conclusions in Sec.~\ref{sec:summary}.

\begin{figure}[t]
	\begin{centering}
		\includegraphics[width=1\linewidth]{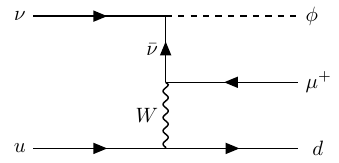}
		\par\end{centering}
	\caption{Feynman diagram for the neutrinophilic scalar radiation via neutrino bremsstrahlung.} 
	\label{fig:splitting}
\end{figure}

\section{Cross sections with a neutrinophilic mediator}
\label{sec:model}
As a benchmark model, we consider a massive scalar $\phi$ that only couples to the SM neutrinos. The effective Lagrangian is given by 
\begin{equation}
\mathcal{L}\supset \frac{1}{2}\lambda_{\alpha\beta}\overline{\nu^c_{\alpha}}\nu_{\beta}\phi+h.c.\,,
\label{eq:scalarL}
\end{equation}
where $\alpha, \beta=e, \mu, \tau$ denotes flavor indices. The scalar $\phi$ can arise from ultraviolet complete models that contain a lepton-number-charged scalar (LeNCS)~\cite{Berryman:2018ogk} or the Majoron~\cite{Gelmini:1980re, Chikashige:1980ui}.

As shown in Fig.\ref{fig:splitting}, the presence of a neutrinophilic scalar will lead to an initial state radiation of $\phi$ when neutrinos scatter off in the detector via the SM CC process. The radiation of $\phi$ will not only produce a wrong sign for the final leptons but also modify the energy spectrum due to the missing energies in the event reconstruction.

To calculate the cross section of the 2-to-3 process $\nu u \to \phi d\mu^+$, we factorize the full process into a neutrino splitting process $\nu (p)\to\bar\nu(k)\phi(q)$ and a neutrino scattering process $\bar\nu u \to d\mu^+$. In the limit $p_T, m_{\phi} \ll E_\nu$, the four-momentum can be written as
\begin{align}
p &= \{E_\nu,0,0,E_\nu\}\,,\\
k &= \{(1-z)E_\nu + \frac{p_T^2}{2(1-z)E_\nu},-p_T,0,(1-z)E_\nu\}\,,\\
q &= \{zE_\nu + \frac{p_T^2+m_\phi^2}{2zE_\nu},p_T,0,zE_\nu\},
\end{align}
and the cross section can be evaluated with
\begin{widetext}
\begin{align}
\sigma_{\nu u \to \phi d\mu^+} & \simeq  \int dzdp_T^2\frac{1}{16\pi^2z}|M_{\nu\to \bar\nu\phi}|^2\left[\frac{1}{(p-q)^2}\right]^2(1-z)\sigma_{\bar\nu u \to d\mu^+}(\hat s)\,,
\label{eq:sigmaXY}
\end{align}
\end{widetext}
where $z$ is the ratio of momentum in z-direction between $\phi$ and the initial neutrino. $p_T$ is the transverse momentum of $\phi$, and $(p-q)^2 = -[p_T^2+(1-z)m_{\phi}^2]/z$ is the momentum transferred of virtual (anti)neutrino. The center-of-mass energy in the hard process $\hat s = (1-z) s$ up to the corrections in the order of $p_T^2/E_\nu^2$. 
In the approximation $p_T, m_{\phi} \ll E_\nu$, the squared amplitudes are
\begin{align}
|M_{\nu\to\bar\nu\phi}|^2 &= \lambda^2 \frac{p^2_T}{1-z}\,.
\end{align}

\begin{figure}[t]
	\begin{centering}
		\includegraphics[width=1\linewidth]{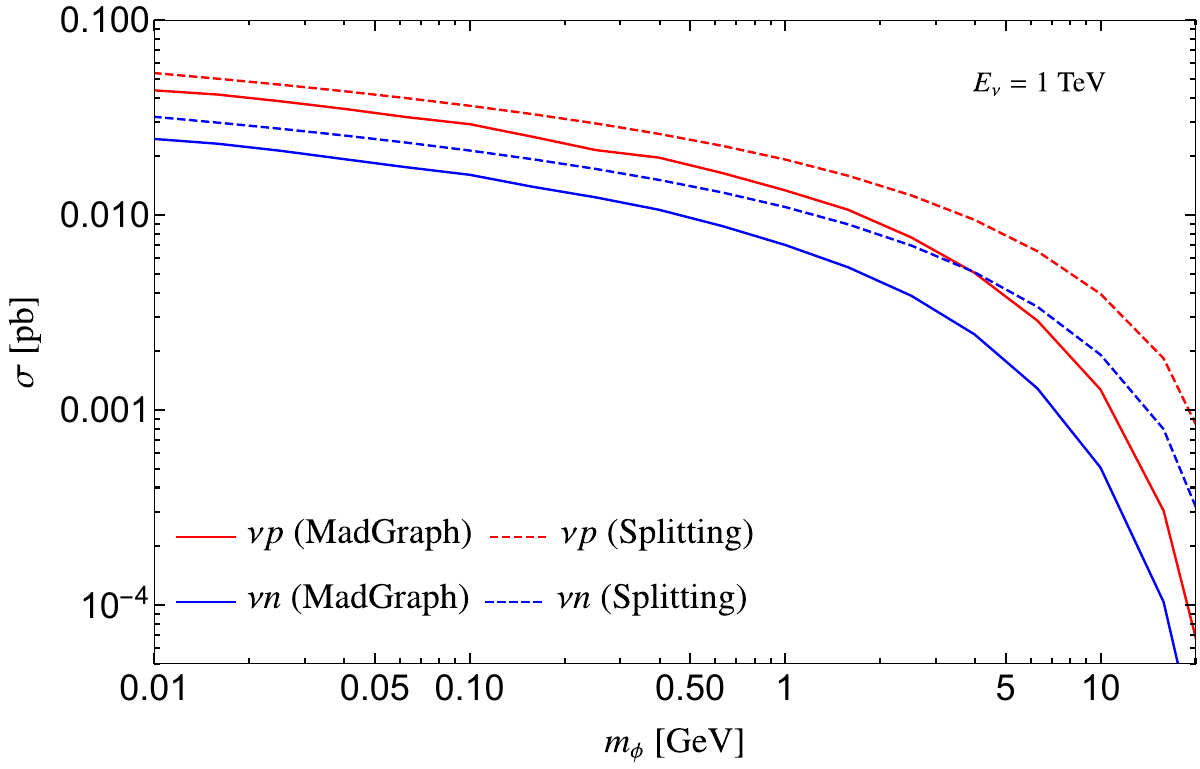}
		\par\end{centering}
	\caption{The cross section of $\phi$ radiation $\sigma_{\nu N\to\phi\mu^+X}$ as a function of $m_\phi$ in the $\nu p$ (red) and $\nu n$ (blue) scattering obtained from the neutrino splitting approximation (dashed) and simulations from MadGraph (solid). Here, we assume $E_\nu= 1$ TeV.} 
	\label{fig:compare}
\end{figure}

For the charged-current process, $\sigma_{\nu \bar u \to \bar d \mu^-}=\sigma_{\bar\nu u \to d \mu^+}= \sigma_{\nu d \to u \mu^-}/3 = \sigma_{\bar\nu \bar d \to \bar u \mu^+}/3 = G_F^2 \hat s /(3\pi)$. The differential cross section can be written as
\begin{align}
\frac{d\sigma_{\nu_\mu u\to\phi d\mu^+}}{dz dp_T^2}&= \frac{G_F^2s}{3\pi}\lambda^2\frac{z(1-z)}{16\pi^2}\frac{p_T^2}{[p_T^2+m_\phi^2(1-z)]^2}.
\end{align} 
After integrating over $z$ from 0 to 1, the differential cross section of the 2-to-3 process to the transverse momentum $p_T$ becomes~\footnote{Note that our result differs by a factor of 9 compared to Eq.~(6) in Ref.~\cite{Kelly:2021mcd}.}
\begin{widetext}
\beqa
\frac{d\sigma_{\nu u\to\phi d \mu^+}}{dp_T}\simeq\frac{G_F^2s}{3\pi}\frac{\lambda^2p_T^3}{8\pi^2m_{\phi}^4} \left[\left(1+\frac{2p_T^2}{m_\phi^2}\right)\text{log}\left(1+\frac{m_\phi^2}{p_T^2}\right)-2\right]\,.
\eeqa
\end{widetext}
To check the analytic results, we also compare the total cross sections at the hadron level with those obtained from the MadGraph simulation.
The total cross section of neutrino bremsstrahlung at the hadron level is given by
\begin{align}
\sigma_{\nu N\to\phi\mu^+X} =\sigma_{\nu u\to\phi d\mu^+} \langle x u_N\rangle + \sigma_{\nu \bar d\to\phi\bar u\mu^+} \langle x \bar d_N\rangle,
\end{align}
where $\langle x q_N\rangle \equiv \int_0^1  x q_N(x)dx$ and $\langle x \bar q_N\rangle \equiv \int_0^1  x \bar q_N(x)dx$ denote the fraction of nucleon momentum carried by quark $u$ and antiquark $\bar d$. Here we used the CT10 PDFs~\cite{Lai:2010vv} and the Mathematica package \texttt{ManeParse}~\cite{Clark:2016jgm} to calculate the cross sections.
In Fig.~\ref{fig:compare}, we compare our analytical results using the neutrino splitting function (dashed lines) with the the ones using \texttt{MadGraph5\_aMC@NLO}~\cite{Alwall:2011uj} and \texttt{FeynRules}~\cite{Christensen:2008py} (solid lines), fixing the neutrino energy at 1 TeV and showing the $\nu p$ (red lines) and $\nu n$ (blue lines) scattering cross section as a function of scalar mass $m_\phi$. From Fig.~\ref{fig:compare}, we see that the cross sections match well in the low-mass region.

We also consider a benchmark model with a vector boson $Z^\prime$ that couples to SM neutrinos dominantly, and the effective Lagrangian is given by 
\begin{equation}
\mathcal{L}\supset g_{\alpha\beta}\bar{\nu}_{\alpha} \gamma^\mu \nu_{\beta}Z_{\mu}^\prime\,,
\label{eq:scalarV}
\end{equation}
where the neutrinophilic boson $Z^\prime$ can arise from a model that contains a new heavy fermion charged under an extra U(1) gauge symmetry and mixed with active neutrinos~\cite{Farzan:2016wym,Bahraminasr:2020ssz}. Since vector interactions conserve the helicity of neutrinos, neutrinos and antineutrinos are not flipped. The calculation of the cross section of the vector boson radiation is the same as Eq.~(\ref{eq:sigmaXY}) except for replacing the squared amplitudes by
\beqa
|M_{\nu\to\nu Z^{\prime}}|^2 &= g^{\prime 2}\left[\frac{p_T^2}{z^2}\left(1-z+\frac{1}{1-z}\right)+\frac{(p_T^2-m_{Z^\prime}^2(1-z))^2}{2m_{Z^\prime}^2(1-z)z^2}\right].\nonumber\\
\eeqa
Therefore, the differential cross section for the vector boson radiation is
\begin{widetext}
\begin{align}
\frac{d\sigma_{\nu_\mu d \to Z^\prime u \mu^-}}{dz dp_T^2}&=\frac{G_F^2s}{\pi}g^{\prime 2}\frac{(1-z)}{32\pi^2z}\frac{p_T^4/m_{Z^\prime}^2+m_{Z^\prime}^2(1-z)^2+2p_T^2(1-z+z^2)}{[p_T^2+m_{Z^\prime}^2(1-z)]^2}.
\end{align} 
\end{widetext}
Note that for the $Z^\prime$ case, there is an infrared divergence and the approximation of squared amplitude cannot be held for $z\to 0$.  

In addition, the cross section of the pseudoscalar (axial vector) case is similar to the scalar (vector) case. This is because in the massless limit of SM neutrinos, we have $\bar u_\nu \Gamma u_\nu + \bar u_\nu \Gamma\gamma^5 u_\nu  = \bar u_\nu \Gamma (1+\gamma^5) u_\nu = 0$, which does not depend on the Lorentz structure of $\Gamma$~\cite{Tsai:1986tx}. 

\section{Analysis of the FASER$\nu$ data}
\label{sec:analysis}
FASER Collaboration has reported the first direct observation of collider neutrinos by using the active electronic components of the FASER detector in 2023~\cite{FASER:2023zcr}. Their dataset contains both the energy distribution and charge identification of the outgoing muons. In this section, we perform an analysis on the FASER$\nu$ 2023 data to put constraints on the neutrinophilic scalar and vector mediators. Constraints on a pseudoscalar (axial vector) mediator are close to those on the scalar (vector) mediator due to similar cross sections in the neutrino massless limit.
Also, since muon neutrino is the dominant component in the forward neutrino fluxes at FASER~\cite{Kling:2021gos}, we only consider the muon flavors for simplicity hereinafter. 

In our analysis, we use MadGraph to generate the neutrino-nucleon interaction cross sections $\sigma_{\nu p}(E_{\nu})$, $\sigma_{\nu n}(E_{\nu})$,
$\sigma_{\bar{\nu}p}(E_{\nu})$ and $\sigma_{\bar{\nu}n}(E_{\nu})$
for $10<E_{\nu}/\textrm{GeV}<10^{4}$. The 4-momenta of the final state leptons and quarks for the SM and the NP cases are also simulated by MadGraph. After neglecting the nuclear effect for high energy neutrinos in the deep inelastic scattering (DIS) region, the neutrino-nucleus interaction cross section $\sigma_{\nu A}$ can be written as
\begin{equation}
\sigma_{\nu A}(E_{\nu})=Z\sigma_{\nu p}(E_{\nu})+(A-Z)\sigma_{\nu n}(E_{\nu})\,,   
\end{equation}
for target nucleus with mass number $A$ and atomic number $Z$. 
Due to the absorption effect during the neutrino propagation in the detector, the neutrino flux decreases  with the traversed depth $X$, and we can get
\begin{equation}
\phi_{\nu}(E_{\nu},X)=\phi_{\nu}(E_{\nu},0)e^{-X/\lambda_{\nu}}\,,   
\label{eq:phinu}    
\end{equation}
where $X\equiv\rho_{d}l$ with $\rho_{d}$ the mass density of the detector and $l$ the neutrino travel distance in the detector. 
The interaction length $\lambda_{\nu}$ here is defined as $\lambda_{\nu}=m_{A}/\sigma_{\nu A}(E_{\nu})$
with $m_{A}$ being the mass of the target nucleus involved. $\phi_{\nu}(E_{\nu},0)$ is the neutrino flux in front of the detector. Given the neutrino-nucleus interaction cross section and the neutrino flux, the number of events per unit neutrino energy per unit muon energy can be calculated by
\begin{align}
\frac{dN}{dE_{\nu}dE_{\mu}} & =\frac{1}{m_{A}}\int dS^{'}dX^{'}dt^{'}\frac{d\sigma_{\nu A}}{dE_{\mu}}\phi_{\nu}(E_{\nu},0)e^{-X'/\lambda_{\nu}}\,.
\label{eq:event_number}
\end{align}
Here the cross sectional area $S$, the depth $X$ of the detector and the data taking period $t$ is integrated over. The differential cross section of neutrino-nucleus interaction $d\sigma_{\nu A}/dE_{\mu}$  is obtained from the MadGraph simulation. 

The neutrino spectra at FASER$\nu$ have been simulated in Ref.~\cite{Kling:2021gos}. 
However, in Ref.~\cite{Kling:2021gos}, the neutrino spectra were computed for a detector with a mass of 1.2 tonnes and the cross sectional area of 25 cm $\times$ 25 cm for LHC Run 3 with the $pp$ collision center of mass energy $\sqrt{s}=13$ TeV and the integrated luminosity $\mathcal{L}_{\rm{int}}=150$ fb$^{-1}$.
Note that the FASER$\nu$ 2023 data were collected at $\sqrt{s}=13.6$ TeV and $\mathcal{L}_{\rm{int}}=35.4$ fb$^{-1}$ with a detector that has a width of 25 cm and a height of 30 cm, and the total mass of the detector is about 1.1 metric tons~\cite{FASER:2023zcr}. Since these settings are different from those assumed in the simulation of Ref. \cite{Kling:2021gos}, in order to calculate the number of events for the FASER$\nu$ 2023 data, we assume that the number of neutrinos passing through the cross sectional area of a certain detector $d\mathcal{N}/dE_{\nu} \approx\phi_{\nu}(E_{\nu},0)St$ and the integrated luminosity in $pp$ collision $\mathcal{L}_{\rm{int}}\approx\mathcal{L}t$, where $\mathcal{L}$ is the corresponding luminosity during the data taking period $t$. Thus, the neutrino energy spectra for the FASER$\nu$ 2023 data can be written as 
\begin{equation}
\bigg(\frac{d\mathcal{N}}{dE_{\nu}}\bigg)_0 \approx \bigg(\frac{d\mathcal{N}}{dE_{\nu}}\bigg) \frac{S_{0}}{S}\frac{\mathcal{L}_{\textrm{int}0}}{\mathcal{L}_{\textrm{int}}}\,,   
\label{eq:Phinu0}
\end{equation}
where $S_{0}$ and $\mathcal{L}_{\textrm{int}0}$ are the cross sectional area of the detector and the $pp$ collision integrated
luminosity for the FASER$\nu$ 2023 data, respectively.  Here $d\mathcal{N}/dE_{\nu}$, $S$ and $\mathcal{L}_{\textrm{int}}$ are the corresponding quantities used in the FASER$\nu$ simulation of Ref. \cite{Kling:2021gos}. Similarly, muon spectra for the proposed experiments FASER$\nu$2 and FLArE at the HL-LHC era with $\sqrt{s}=14$ TeV and $\mathcal{L}_{\rm{int}}=3000$ fb$^{-1}$ can be also computed using the detector configuration listed in Table \ref{tab:detector}. 

\begin{table}
	\begin{centering}
		\begin{tabular}{|c|c|c|c|}
			\hline 
			Detector & Mass {[}tonne{]} & $S$ {[}cm$^{2}${]} & $\mathcal{L}_{\textrm{int}}$ {[}fb$^{-1}${]} \tabularnewline
			\hline 
			FASER$\nu$~\cite{Kling:2021gos} & 1.2 & $25\times25$ & 150  \tabularnewline
			\hline 
			FASER$\nu$ 2023~\cite{FASER:2023zcr} & 1.1 & $25\times30$ & 35.4 
			\tabularnewline
			\hline 
			FLArE~\cite{Feng:2022inv} & 10 & $100\times100$ & 3000 
			\tabularnewline
			\hline
			FASER$\nu$2~\cite{Feng:2022inv} & 20 & $40\times40$ & 3000 
			\tabularnewline
			\hline 
		\end{tabular}
	\end{centering}
	\caption{Configuration of neutrino detectors considered in this paper at LHC run 3 and the HL-LHC era. 
	}
	\label{tab:detector}
\end{table}

\begin{figure}
	\begin{centering}
		\includegraphics[width=1\linewidth]{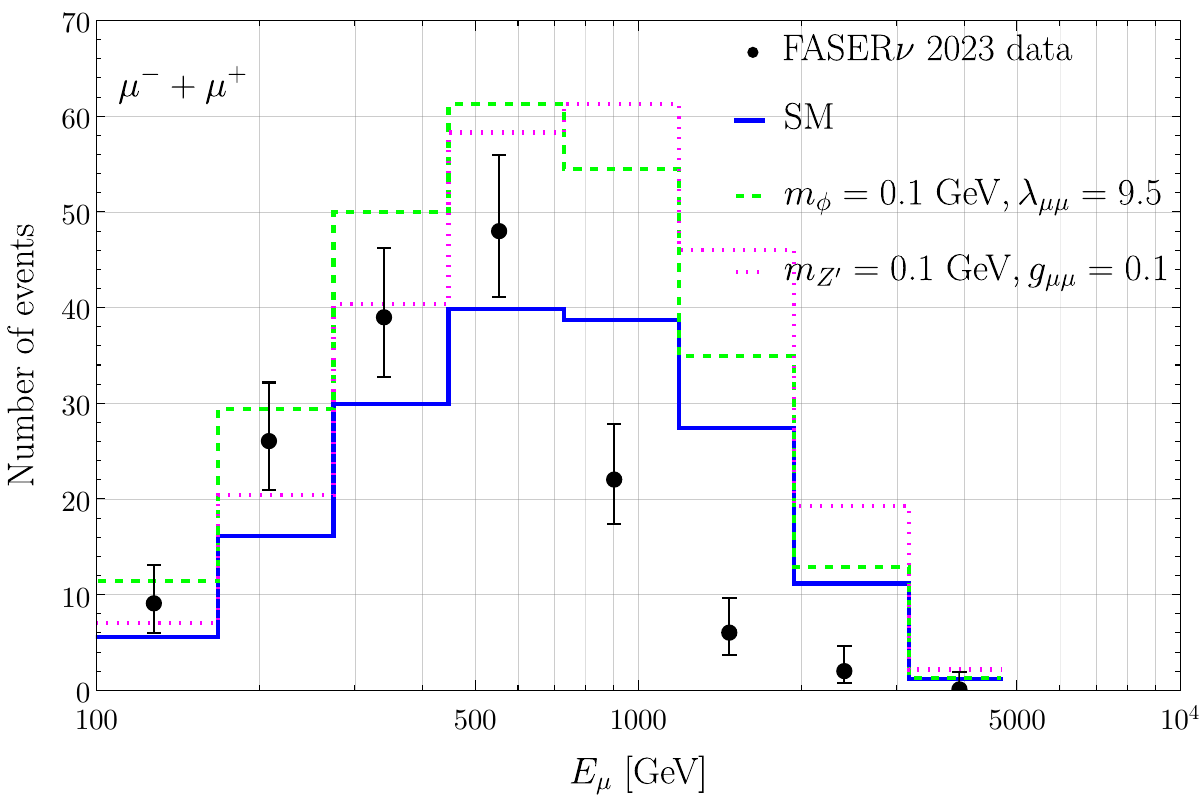}
		\includegraphics[width=1\linewidth]{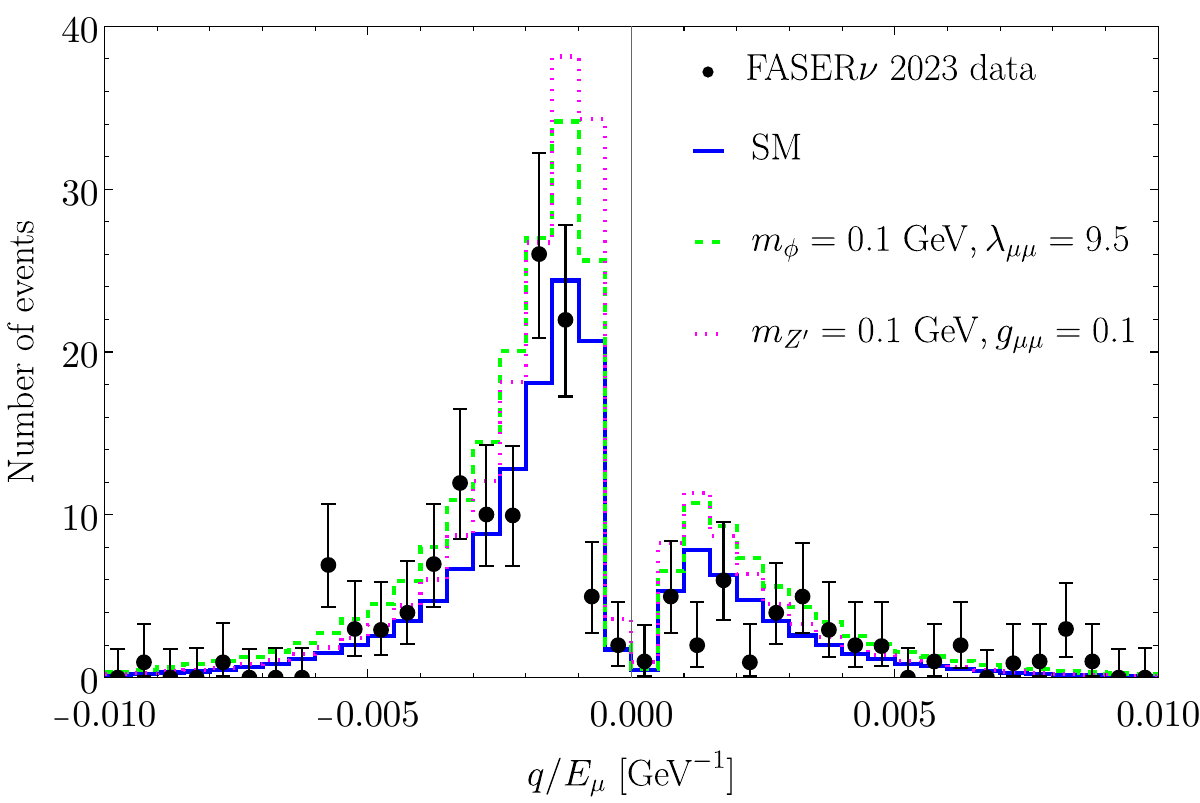}
		\par\end{centering}
	\caption{The measured and predicted number of events as a function of the muon energy $E_{\mu}$ (left panel) and $q/E_{\mu}$ (right panel) for the SM and NP cases. The SM case is shown as the blue dashed line, and the green dashed (magenta dotted) line corresponds to the NP case with $m_{\phi}=0.1$ GeV, $\lambda_{\mu\mu}=9.5$ ($m_{Z'}=0.1$ GeV, $g_{\mu\mu}=0.1$). The data points are taken from the FASER$\nu$ measurements~\cite{FASER:2023zcr}. 
		\label{fig:Fasernu2023-SM-Emu-q2Emu}}
\end{figure}

After taking into account these approximations, the number of events per unit neutrino energy per unit muon energy for the  FASER$\nu$ 2023 data can be written as
\begin{equation}
\frac{dN}{dE_{\nu}dE_{\mu}} \approx\frac{1}{m_{A}} \bigg(\frac{d\mathcal{N}}{dE_{\nu}}\bigg)_0 \left(\frac{1}{S_{0}}\int dS^{'}dX^{'}\frac{d\sigma_{\nu A}}{dE_{\mu}}e^{-X'/\lambda_{\nu}}\right)\,.
\label{eq:nEvt_FASERnu2023}
\end{equation}  
Since only muons produced in CC interactions that traverse the entire length of the FASER detector were collected, we apply the following event selection criteria in our simulation of the data: 
\begin{itemize}
	\item The polar angle $\theta$ of the reconstructed track is required to satisfy $\theta<25$ mrad;
	\item The reconstructed track's extrapolation to the FASER$\nu$ scintillator must be at a distance of $r_{\textrm{veto}\nu}<120$ mm from the FASER$\nu$ scintillator center;  
	\item The reconstructed track's extrapolation to the interface tracking station must lie within $95$ mm of the detector's central axis;
	\item The reconstructed track traverses the three tracking spectrometer stations with each having an effective diameter of 200 mm.
\end{itemize}
These cuts are applied in our simulation of the number of events when performing the integration over the cross sectional area $S$ and the traversed depth $X$ of Eq. (\ref{eq:nEvt_FASERnu2023}). The measured and predicted number of events as a function of the muon energy $E_{\mu}$ for the SM case are shown in the left panel of Fig. \ref{fig:Fasernu2023-SM-Emu-q2Emu}. 
Also, the muon neutrino events in the FASER$\nu$ 2023 dataset were measured by the active electronic components of the FASER detector, which have the ability to reconstruct the muon charge~\cite{FASER:2023zcr,FASER:2022hcn}. 
A similar plot with the muon
charge identification, i.e., the number of events as a function of $q/E_{\mu}$ (where $q$ is the charge of the muon), is shown in the right panel of Fig.~\ref{fig:Fasernu2023-SM-Emu-q2Emu}. 
In Fig. \ref{fig:Fasernu2023-SM-Emu-q2Emu}, we also show the muon spectra of a neutrinophilic scalar and vector mediator for illustration.
We take $m_{\phi}=0.1$ GeV, $\lambda_{\mu\mu}=9.5$ for the scalar case and $m_{Z'}=0.1$ GeV, $g_{\mu\mu}=0.1$ for the vector case in the plots to show how the inclusion of the NP can affect the distributions of $E_{\mu}$ and $q/E_{\mu}$. It can be seen that NP with the vector mediator changes the number of events distributions more significantly than that with the scalar mediator.

We use the dataset of the number of events as a function of $q/E_{\mu}$ in our analysis since it contains additional information with the charge identification.
To evaluate the statistical significance of the
NP with the neutrinophilic mediators, we define 
\begin{equation}
\chi^{2}=\sum_{i}2(\alpha N_{i}^{th}-N_{i}^{exp}+N_{i}^{exp}\ln\frac{N_{i}^{exp}}{\alpha N_{i}^{th}})+\frac{(1-\alpha)^{2}}{\sigma_{\alpha}^{2}},    
\end{equation}
where $\sigma_{\alpha}=10\%$ is the percent uncertainty in
the neutrino flux normalization~\cite{Kling:2021gos}, $N_{i}^{exp}$ is the number of events in the $i$th bin of the FASER$\nu$ 2023 with charge identification, and $N_{i}^{th}$
is the theoretical prediction for the number of events in the $i$th
bin. The best fit of the SM predictions yields $\chi^2_{\rm{min}}/\rm{d.o.f.}=80.5/40$ with $\alpha=1.07$. We find that the 90$\%$ C.L. upper bounds for the scalar mediator (e.g., $\lambda_{\mu\mu}=13.1@m_{\phi}=1$ GeV) are much weaker than the existing bounds from the kaon and the $Z$ boson invisible decays \cite{Kelly:2021mcd, Brdar:2020nbj}. The bounds on the vector mediator $Z'$ is much more stringent. From the right panel of Fig. \ref{fig:Constraints-on-coupling}, we can see that bounds on the vector mediator are comparable to the existing bounds \cite{Bahraminasr:2020ssz,Laha:2013xua, Bakhti:2017jhm} at $m_{Z^\prime}\approx 0.2$ GeV. 

\section{Sensitivities at the FLArE and FASER$\nu$2 experiments}
\label{sec:simulation}
Since the constraints on the parameter space of the neutrinophilic mediators are not strong from the FASER$\nu$ 2023 data, we turn to the next stage of the FPF program and study the sensitivities to neutrinophilic mediators at FLArE and FASER$\nu$2. The experiments FASER2 and FASER$\nu$2 are the upgraded versions of FASER and FASER$\nu$, respectively. The ideal location of the FASER$\nu$2 detector is still in the front of the FASER2 spectrometer along the beam collision axis. The total volume of the tungsten target of the FASER$\nu$2 detector is $40$ cm $\times$ $40$ cm $\times$ $6.6$ m with the mass of $20$ tonnes~\cite{Feng:2022inv}. The experiment FLArE is also able to measure millions of neutrino interactions and identify the neutrino types. A detector with a fiducial mass of approximately 10 tonnes and a cross sectional area of $100$ cm $\times$ $100$ cm is envisioned for FLArE~\cite{Feng:2022inv}. 

Since the neutrinophilic mediator will decay dominantly into neutrinos, it appears invisible
after production in neutrino beamstrahlung. The resulting missing
transverse momentum $\slashed{p}_T$ can be reconstructed from the transverse momenta of the final state muon and quark, i.e., $\slashed{p}_T=\left|\vec{p}_{T\mu}+\vec{p}_{Tq}\right|$. Since the double distributions of the number of events contain more information than a single distribution and generally perform better in a sensitivity analysis, we choose the transverse momentum of the final state muon $p_{T\mu}$ as the second kinematic observable in addition to $\slashed{p}_T$. We also take into account of the effects of a finite energy resolution for the detectors, which is approximated by smearing on the muon and quark energies. We assume that the muon energy resolution and the hadronic energy resolution is $5\%$ and $15\%$~\cite{Kelly:2021mcd}, respectively. 

\begin{figure}[t]
	\begin{centering}
		\includegraphics[width=1\linewidth]{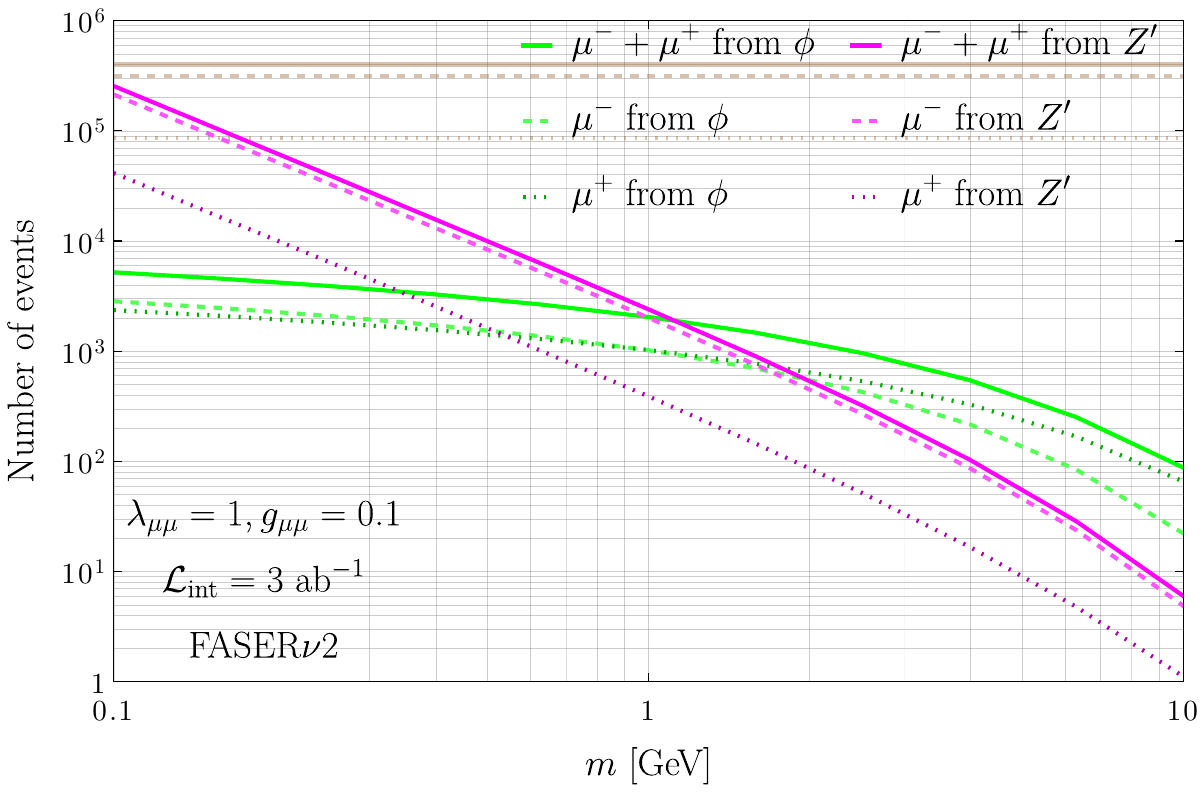}
		\par\end{centering}
	\caption{Total number of events as a function of the mass of the neutrinophilic mediators at FASER$\nu$2. The green and magenta lines corresponds to the $\phi$ and $Z'$ cases with $\lambda_{\mu\mu}=1$ and $g_{\mu\mu}=0.1$, respectively. The SM results are shown by brown curves.} 
	\label{fig:nEvt_m_f}
\end{figure}

After the integration of Eq.~(\ref{eq:event_number}) over the cross sectional area, the depth of the detector and the data taking period, the number of events with respect to the missing transverse momentum and muon transverse momentum can be written as
\begin{equation}
\frac{dN}{dE_{\nu}d\slashed{p}_Tdp_{T\mu}} =\frac{X_d}{m_{A}} 
\frac{d\mathcal{N}}{dE_{\nu}} \frac{d^2\sigma_{\nu A}}{d\slashed{p}_Tdp_{T\mu}},    
\end{equation}
where $X_d$ is the overall traversed depth of the detector defined below Eq.~(\ref{eq:phinu}). Note that for the future experiments, we do not apply the cuts that are dependent on the positions of neutrinos in the detector in selecting events as we do in analyzing the experimental data in Sect. \ref{sec:analysis}. The neutrino-nucleus double differential cross section $d\sigma_{\nu A}/d\slashed{p}_Tdp_{T\mu}$ is also obtained from the simulation of MadGraph. In Fig. \ref{fig:nEvt_m_f}, we show the total number of events as a function of the mediator mass at FASER$\nu$2 for the scalar and vector case with $\lambda_{\mu\mu}=1$ and $g_{\mu\mu}=0.1$, respectively.

To improve the sensitivity to the neutrinophilic mediators at the future FPF experiments, we perform a $\chi^2$ analysis
\begin{equation}
\chi^{2}=\frac{N_{S}^{2}}{N_{B}+(\sigma_{B}N_{B})^{2}}\,,  
\end{equation}
where $N_S$ ($N_B$) is the total number of events of the signal (SM CC background), $\sigma_{B}=10\%$ is percent uncertainty in the SM background~\cite{Kling:2021gos}. To find regions where the NP signal is more prominent in the $($$\slashed{p}_T$$,p_{T\mu})$ plane, we also calculate the ratio of $N_S/(N_S+N_B)$ as in Ref. \cite{Kelly:2021mcd}, and perform the cuts by requiring $p_{T\mu}\lesssim4\slashed{p}_T/3$ ($p_{T\mu}\lesssim1.9\slashed{p}_T$) in the $($$\slashed{p}_T$$,p_{T\mu})$ plane for the scalar (vector) mediator case. 

\begin{figure}
	\begin{centering}
		\includegraphics[width=1\linewidth]{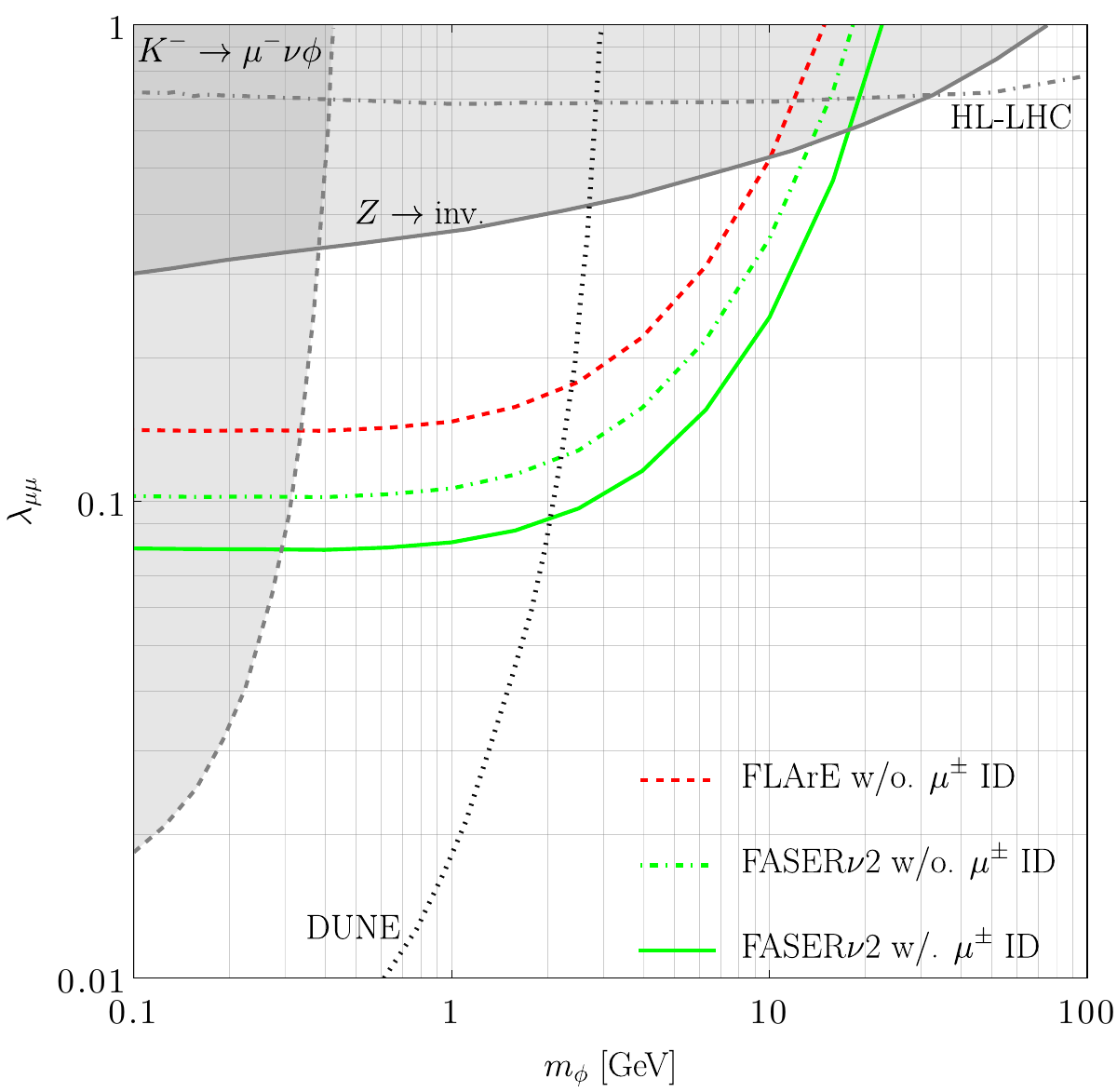}
		\includegraphics[width=1\linewidth]{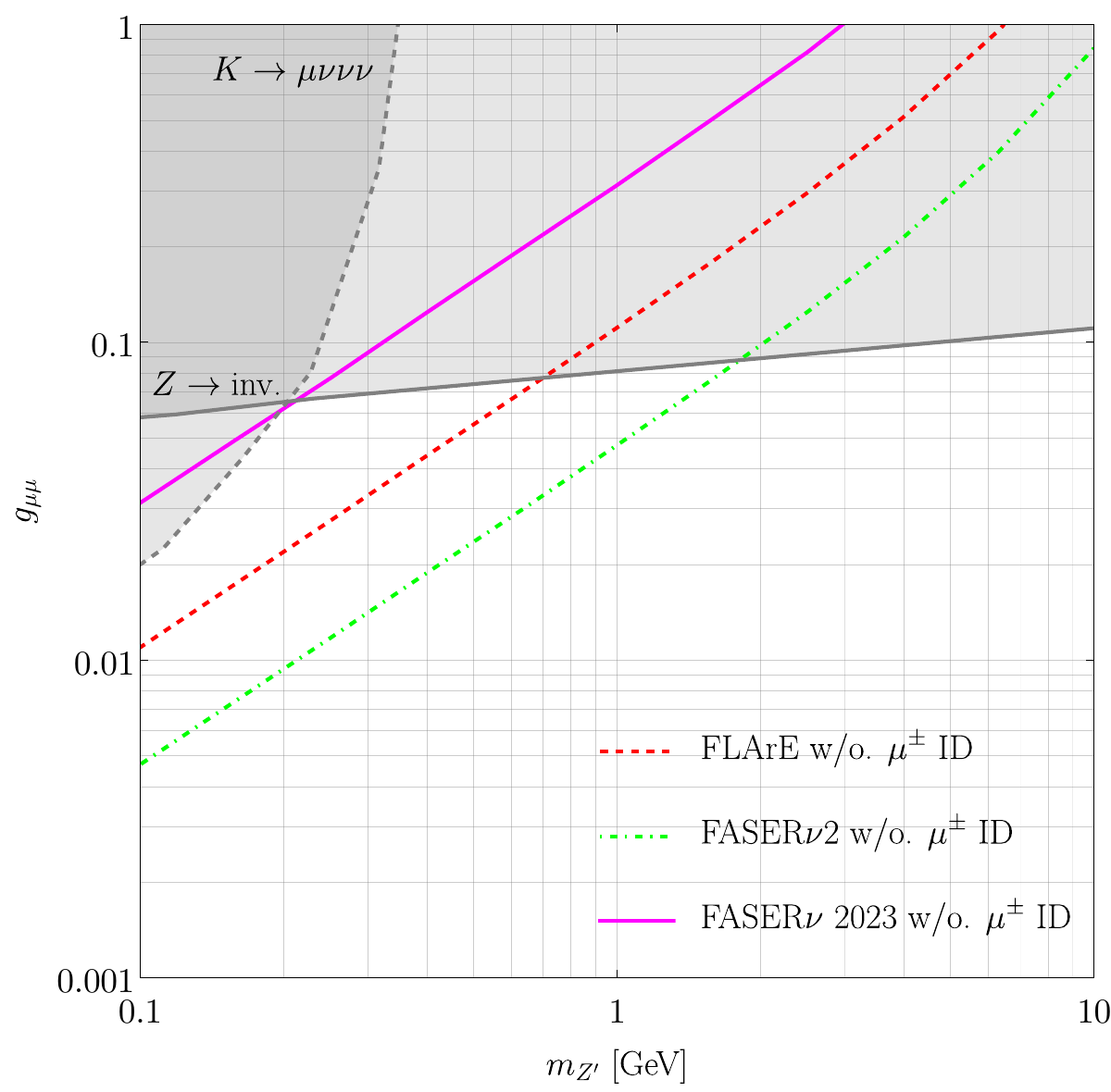}
		\par\end{centering}
	\caption{Sensitivities to a scalar (left panel) or vector (right panel) neutrinophilic mediator. The magenta solid curve shows the constraints from the FASER$\nu$ 2023 data. The red and green curves correspond to the bounds at FLArE and FASER$\nu$2, respectively. The green solid (dot-dashed) curve in the left panel corresponds to the bound at FASER$\nu$2 with (without) the charge identification information. The gray shaded regions represent the regions that are ruled out by the charged kaon and the $Z$-boson invisible decays for the scalar~\cite{Kelly:2021mcd,  Brdar:2020nbj} and vector neutrinophilic mediators~\cite{Bahraminasr:2020ssz,Laha:2013xua, Bakhti:2017jhm}. The DUNE~\cite{Kelly:2019wow} and HL-LHC~\cite{deGouvea:2019qaz} projections are overlaid.  
\label{fig:Constraints-on-coupling}}
\end{figure}

Our results are presented in the left and right panels of Fig.~\ref{fig:Constraints-on-coupling} for the scalar and vector neutrinophilic mediators, respectively. The parameter space that are ruled out by the charged kaon and the $Z$-boson invisible decays are shown as the gray shaded regions in the left and right panels for the scalar~\cite{Kelly:2021mcd,  Brdar:2020nbj} and vector neutrinophilic mediators~\cite{Bahraminasr:2020ssz,Laha:2013xua, Bakhti:2017jhm}, respectively. The constraints at FLArE (FASER$\nu$2) are shown as the red (green) curves.
From the left panel of Fig.~\ref{fig:Constraints-on-coupling}, we see that both FLArE and FASER$\nu$2 can impose stronger bounds on the scalar neutrinophilic mediator than the existing bounds. The constraints at FASER$\nu$2 are stronger than those at FLArE. In particular, since the radiation of a scalar will produce a wrong sign charged lepton, we also take into account the charge identification information in the analysis for the scalar mediator.
From the left panel of Fig.~\ref{fig:Constraints-on-coupling}, we see that the constraints on the scalar mediator can be improved if the charge identification information is provided. The bounds can reach 0.08 (0.1) for $m_\phi\lesssim1$ GeV with (without) charge identification at FASER$\nu$2. For the vector case, from the right panel of Fig.~\ref{fig:Constraints-on-coupling}, we see that FASER$\nu$2 also yields a stronger bound than FLArE, and the bounds at FLArE (FASER$\nu$2) can be more stringent than the existing ones for $m_{Z^\prime}$ below 0.7 (1.8) GeV.
 
\section{Summary}
\label{sec:summary}
The FASER$\nu$ experiment has made the first observation of collider neutrinos using the active electronic components of the FASER detector in 2023. The measured spectrum of high-energy neutrinos scattering off detector nucleons are sensitive to neutrinophilic mediators with GeV-scale masses. We study the sensitivities of current and future forward neutrino experiments to the neutrinophilic mediator.
We find that constraints on a pseudoscalar (axial vector) mediator are close to those on the scalar (vector) mediator since they have similar cross sections in the massless neutrino limit.
We have performed an analysis of the current FASER$\nu$ 2023 data, and find that the bounds on the scalar neutrinophilic mediator from the current FASER$\nu$ data are much weaker than the existing bounds, and the bounds on the vector neutrinophilic mediator from the current data are comparable to the existing bounds at $m_{Z^\prime}\approx 0.2$ GeV. We also study the sensitivities of future FPF experiment including FLArE and FASER$\nu$2 to a neutrinophilic scalar or vector mediator by using both the missing transverse momentum and the charge identification information.  We find that FLArE and FASER$\nu$2 can impose stronger bounds on both the scalar and vector neutrinophilic mediators than the existing bounds. The constraints on the scalar mediator can reach 0.08 (0.1) for $m_\phi\lesssim1$ GeV with (without) muon charge identification at FASER$\nu$2.  

\begin{acknowledgments}
We would like to thank Tomoko Ariga, Junmou Chen, and Zhen Hu for useful discussions. W. B. is supported by the National Natural Science Foundation of China under Grant No. 12105376. J.L. is supported by the National Natural Science Foundation of China under Grant Nos.~12275368 and the Fundamental Research Funds for the Central Universities, Sun Yat-Sen University under Grant No. 24qnpy116.
H.L. is supported by Azrieli foundation and by the U.S. Department of Energy under Grant Contract DE-SC0012704.	
\end{acknowledgments}

\twocolumngrid
\vspace{-8pt}
\section*{References}
\vspace{-10pt}
\def\bibsection{}
\bibliography{refs}

\end{document}